\documentclass[10pt]{article}
\usepackage[T1]{fontenc}
\usepackage[a4paper,margin=20mm,columnsep=6mm]{geometry}
\usepackage{lmodern,amsmath,amssymb,graphicx,multirow,dcolumn,bm,xcolor}
\usepackage{cite}
\usepackage{xurl}
\usepackage[hidelinks]{hyperref}
\usepackage{lineno}
\graphicspath{{Images/}}
\newcommand{\beq}{\begin{eqnarray}}
\newcommand{\eeq}{\end{eqnarray}}
\begin{document}
\begin{center}
{\LARGE\bfseries Probing high-momentum components in $^{16}$O via zero-degree $(p,d)$ cross-section ratios\par}\medskip
X.~Wang$^{1,2}$,
H.~J.~Ong$^{3,4,5,2,6,*}$,
S.~Terashima$^{3,1,*}$,
I.~Tanihata$^{2}$,
Y.~K.~Tanaka$^{7}$,
H.~Geissel$^{8,9,\dagger}$,
N.~Aoi$^{2,a}$,
Y.~Ayyad$^{10,11}$,
J.~Benlliure$^{11}$,
F.~Farinon$^{9}$,
H.~Fujioka$^{12}$,
J.~Gellanki$^{13,14}$,
C.~L.~Guo$^{15}$,
E.~Haettner$^{9,13}$,
W.~L.~Hai$^{1}$,
M.~N.~Harakeh$^{9,13}$,
C.~Hornung$^{8}$,
K.~Itahashi$^{6,16}$,
R.~Janik$^{17,\dagger}$,
N.~Kalantar-Nayestanaki$^{13}$,
R.~Knöbel$^{8,9}$,
N.~Kurz$^{9}$,
K.~Miki$^{18}$,
I.~Mukha$^{9}$,
T.~Myo$^{19,2}$,
T.~Nishi$^{6}$,
D.~Y.~Pang$^{1}$,
S.~Pietri$^{9}$,
A.~Prochazka$^{9}$,
C.~Rappold$^{20,9}$,
M.~P.~Reiter$^{8,b}$,
J.~L.~Rodríguez-Sánchez$^{10,c}$,
C.~Scheidenberger$^{9,8,21}$,
H.~Simon$^{9}$,
B.~Sitar$^{17}$,
P.~Strmen$^{17,\dagger}$,
B.~H.~Sun$^{1}$,
K.~Suzuki$^{2}$,
H.~Weick$^{9}$,
J.~S.~Winfield$^{9,\dagger}$,
X.~D~Xu$^{3,4,9}$,
J.~W.~Zhao$^{1}$\par\smallskip
{\itshape Super-FRS Experiment Collaboration\par}\medskip
{\fontsize{8.5}{9.5}\selectfont
\noindent\textsuperscript{1}School of Physics, Beihang University, Beijing 100191, China\par
\noindent\textsuperscript{2}Research Center for Nuclear Physics, The University of Osaka, Ibaraki, Osaka 567-0047, Japan\par
\noindent\textsuperscript{3}State Key Laboratory of Heavy Ion Science and Technology, Institute of Modern Physics, Chinese Academy of Sciences, Lanzhou 730000, China\par
\noindent\textsuperscript{4}School of Nuclear Science and Technology, University of Chinese Academy of Sciences, Beijing 100049, China\par
\noindent\textsuperscript{5}Joint Department for Nuclear Physics, Lanzhou University and Institute of Modern Physics, Chinese Academy of Sciences, Lanzhou 730000, China\par
\noindent\textsuperscript{6}Nishina Center for Accelerator-Based Science, RIKEN, 2-1 Hirosawa, Wako, 351-0198 Saitama, Japan\par
\noindent\textsuperscript{7}High Energy Nuclear Physics Laboratory, RIKEN, 2-1 Hirosawa, Wako, 351-0198 Saitama, Japan\par
\noindent\textsuperscript{8}Universität Giessen, Heinrich-Buff-Ring 16, 35392 Giessen, Germany\par
\noindent\textsuperscript{9}GSI Helmholtzzentrum für Schwerionenforschung GmbH, Planckstraße 1, 64291 Darmstadt, Germany\par
\noindent\textsuperscript{10}Universidade de Santiago de Compostela, 15782 Santiago de Compostela, Spain\par
\noindent\textsuperscript{11}IGFAE, Universidade de Santiago de Compostela, 15782 Santiago de Compostela, Spain\par
\noindent\textsuperscript{12}School of Science, Institute of Science Tokyo, 2-12-1 Ookayama, Meguro, Tokyo 152-8551, Japan\par
\noindent\textsuperscript{13}ESRIG, University of Groningen, Zernikelaan 25, 9747 AA Groningen, The Netherlands\par
\noindent\textsuperscript{14}Department of Physics Hansraj College, University of Delhi, Delhi 110006,  India\par
\noindent\textsuperscript{15}Department of Radiation Oncology, National Cancer Center/National Clinical Research Center for Cancer/Cancer Hospital, Chinese Academy of Medical Sciences and Peking Union Medical College, Beijing 100021, China\par
\noindent\textsuperscript{16}Department of Physics, The University of Osaka, Toyonaka, Osaka 560-0043, Japan\par
\noindent\textsuperscript{17}Comenius University Bratislava, Mlynská dolina, 842 48 Bratislava, Slovakia\par
\noindent\textsuperscript{18}Tohoku University Department of Physics, Graduate School of Science and Faculty of Science Aza-Aoba 6-3, Aramaki, Aoba-ku, Sendai 980-8578, Japan\par
\noindent\textsuperscript{19}General Education, Faculty of Engineering, Osaka Institute of Technology, Osaka 535-8585, Japan\par
\noindent\textsuperscript{20}Instituto de Estructura de la Materia, CSIC, Madrid, Spain\par
\noindent\textsuperscript{21}Helmholtz Forschungsakademie Hessen fuer FAIR (HFHF), GSI Helmholtzzentrum für Schwerionenforschung, Campus Giessen, 35392 Giessen, Germany\par
\noindent\textsuperscript{a}N.~Aoi (Present address: Center for Nuclear Study, University of Tokyo, Tokyo 113-0033, Japan)\par
\noindent\textsuperscript{b}M.~P.~Reiter (Present address: School of Physics and Astronomy, University of Edinburgh, Edinburgh EH9 3FD, United Kingdom)\par
\noindent\textsuperscript{c}J.~L.~Rodríguez-Sánchez (Present address: CITENI, Campus Industrial de Ferrol, Universidade da Coruna, 15403, Ferrol, Spain)\par
\noindent $^{\dagger}$Deceased.\par
\noindent $^{*}$Correspondence: H.~J.~Ong (\href{mailto:onghjin@impcas.ac.cn}{onghjin@impcas.ac.cn}); S.~Terashima (\href{mailto:tera@impcas.ac.cn}{tera@impcas.ac.cn})\par}
\end{center}

\noindent{\small\textbf{Abstract.} The $^{16}$O$(p,d)^{15}$O reaction was measured at zero degrees for incident-proton energies of 403, 604, 907 and 1209 MeV. Cross-section ratios for the 6.18-MeV (3/2$^-$) and 5.24-MeV (5/2$^+$) excited-state transitions, normalized to the ground-state transition, were previously reported in Ref.~\cite{Wang2026}, together with their implications for high-momentum neutron components in $^{16}$O. The present work extends that study by providing additional experimental and data-analysis details, and by reporting for the first time the ratio $R_{7.3}$, defined as the cross section populating the 7.28-MeV (7/2$^+$) state relative to that for the ground state. Both positive-parity state ratios exhibit a pronounced momentum dependence, consistent with representative variational Monte Carlo trends.\par\smallskip \textbf{Keywords.} Nuclear structure $\cdot$ Neutron pickup $\cdot$ Nucleon momentum distributions $\cdot$ Tensor interactions\par\smallskip\textbf{Acknowledgements.} The authors thank the accelerator group and operators, the GSI target laboratory, and the engineers of the FRS group for their support.\par}
\twocolumn

\section{Introduction}
The nuclear force is governed by nucleon-nucleon interactions that include central, spin-orbit, and tensor components. Among these, tensor interactions, arising primarily from pion exchange, play an essential role in nuclear binding and saturation density~\cite{Pieper2001, Bethe1971}. 
{\it Ab initio} Green’s function Monte Carlo calculations for nuclei up to $^{12}$C have confirmed their dominant role in reproducing low-lying energy spectra~\cite{Pieper2001}. 
Rigorous few-body calculations further demonstrate that tensor interactions contribute approximately 80\% of the deuteron binding energy and about 50\% in the alpha particle~\cite{Ikeda2010,Sakai1974},  underscoring their fundamental importance to nuclear structure.

Tensor interactions naturally generate high-momentum nucleons in nuclei by coupling states below the Fermi surface to configurations above it. Theoretical studies have predicted that the resulting high-momentum components are most prominent around 2 fm$^{-1}$ (approximately 400 MeV/$c$)~\cite{Neff2003,Schiavilla2007,Horiuchi2007,Alvioli2008,Lyu2020}. This momentum scale corresponds to the region where the tensor interactions are most effective in promoting $pn$ pairs to relative $D$-states, through mixing of two-particle-two-hole (2p-2h) configurations~\cite{Myo2007,Myo2009}.

Experimental evidence for tensor-induced high-momentum components has emerged from various scattering measurements. Electron-scattering experiments, including ($e$,$e'p$), ($e$,$e'pp$) and ($e$,$e'pn$) reactions, have reported the presence of high-momentum nucleons and strong $pn$ correlations relative to $pp$ pairs~\cite{Benhar1993,Blomqvist1998,Subedi2008,Hen2017,Duer2018,Korover2021}, consistent with theoretical predictions. Further evidence was reported from proton-induced nucleon-knockout reactions~\cite{Mardor1998,Tang2003,Patsyuk2021}.
More detailed spectroscopic information has been obtained from ($p$,$d$) neutron-pickup studies on $^{16}$O at incident energies of 198-392 MeV and scattering angles around 10$^\circ$~\cite{Ong2013}, together with data at lower and higher energies~\cite{Snelgrove1969,Roos1975,Smith1984}. These measurements revealed a distinct difference in the momentum dependence between transitions to negative- and positive-parity final states, which was interpreted as evidence for $sd$-shell admixtures arising from tensor-induced 2p-2h correlations. An exclusive ($p$,$pd$) experiment on $^{16}$O under neutron-pickup kinematics has also been performed, revealing that the cross section depends strongly on the spin and isospin of the populated final states~\cite{Terashima2018}.

\begin{figure}
\includegraphics[width=\columnwidth]{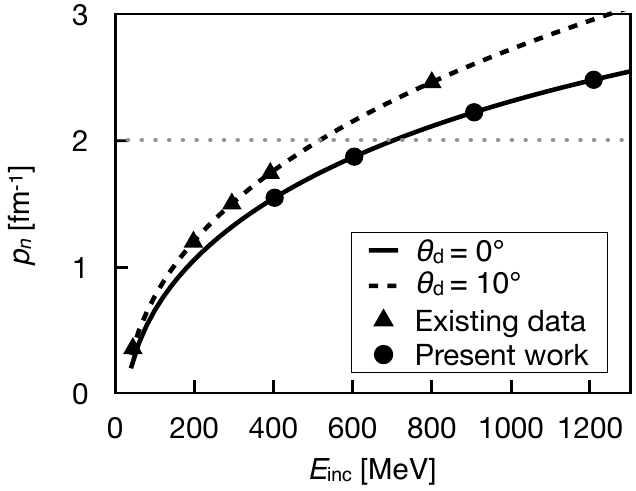}%
\caption{Momenta of the picked-up neutron ($p_{n}$) in the $^{16}$O$(p,d)^{15}$O$_{\textrm{g.s.}}$ reaction as functions of the incident proton energy ($E_{\rm inc}$) for deuteron scattering angles ($\theta_{d}$) of $0^\circ$ and $10^\circ$ in the center-of-mass frame. For reference, selected points with available data at $0^\circ$ and 403--1209 MeV~\cite{Wang2026}, and data at $10^\circ$ and 45~MeV~\cite{Snelgrove1969}, 198--392~MeV~\cite{Ong2013}, and 800~MeV~\cite{Smith1984} are shown.}
\label{fig:q_plot}
\end{figure}

In an earlier Letter~\cite{Wang2026}, we reported $^{16}$O($p$,$d$)$^{15}$O measurements  at zero degrees and at incident proton energies of 403, 604, 907, and 1209 MeV. These energies cover the momentum transfer around 2 fm$^{-1}$ (see Fig.~\ref{fig:q_plot}) where tensor-induced components are expected to be enhanced. The zero-degree geometry minimizes distortion effects and ensures that the cross sections can be directly related to the internal neutron momentum via the kinematic relation $\vec{p}_n = \vec{k}_d - \vec{k}_p$ under the Born approximation~\cite{Chew1950}, where $\vec{k}_p$ and $\vec{k}_d$ are the momenta of the incident proton and the emitted deuteron, respectively. While part of the present results -- specifically the momentum dependence of the ratios $R_{6.2}$ (6.18-MeV $3/2^-$ state) and $R_{5.2}$ (5.24-MeV $5/2^+$ state) relative to the ground state -- has been reported previously~\cite{Wang2026}, this article provides a complete account of the experimental procedure and data analysis, and reports for the first time the additional ratio $R_{7.3}$ for the 7.28-MeV $7/2^+$ state.

The paper is organized as follows. Section~\ref{sec:experiment} describes the experimental setup. Section~\ref{sec:data_analysis} presents the analysis method, including particle identification, momentum reconstruction,  reconstruction of excitation-energy spectra, and determination of cross-section ratios. Section~\ref{sec:results_and_discussion} reports the experimental results and discussion, with emphasis on the momentum dependence of the measured ratios and the new $R_{7.3}$ result. Finally, a summary is given in Section~\ref{sec:summary}.

\section{Experiment} 
\label{sec:experiment}

\begin{figure}
\includegraphics[width=\columnwidth]{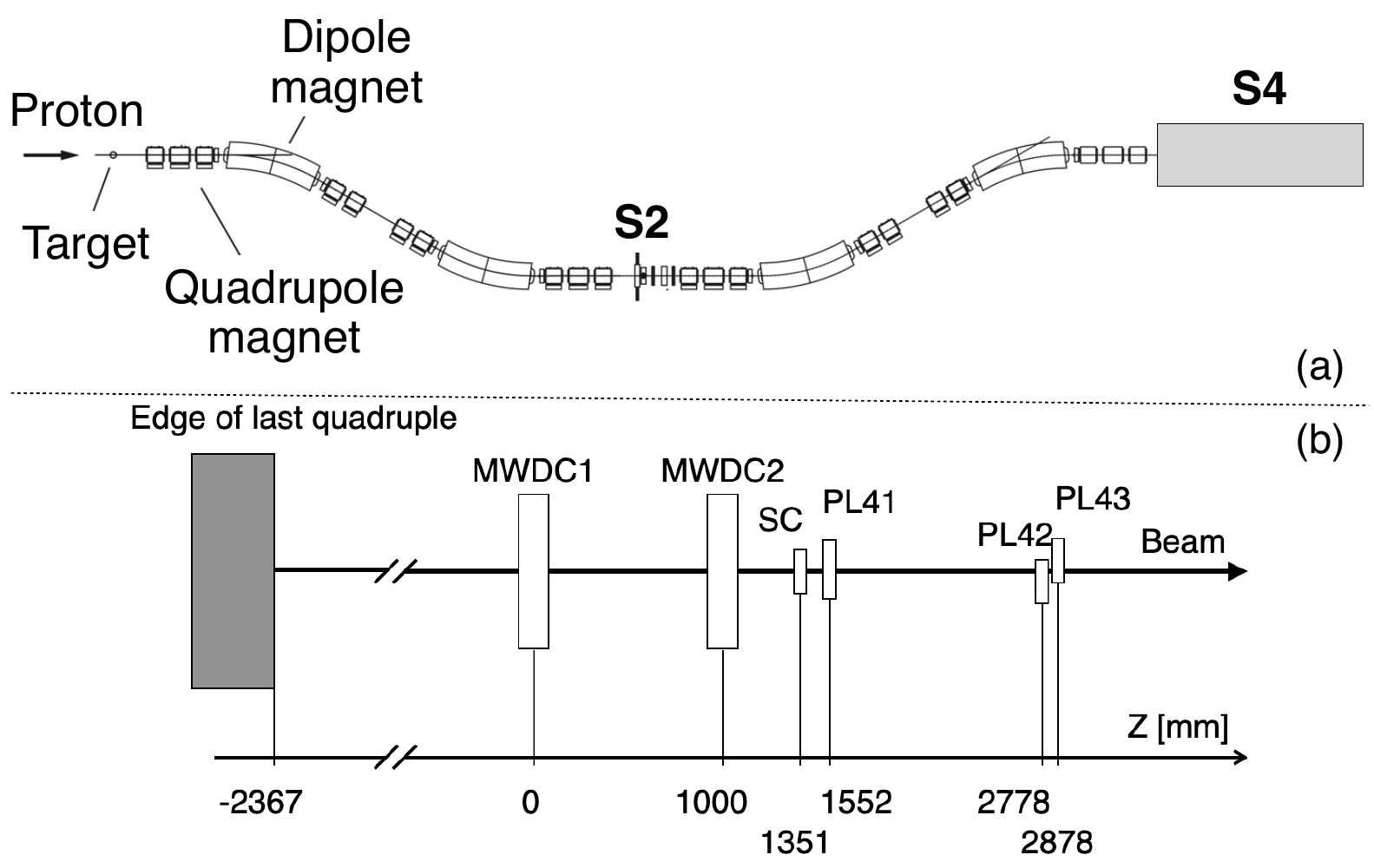}%
\caption{Experimental setup at the FRS (a) and S4 (b). Deuterons emitted from the target were momentum-analyzed by the FRS and detected at the dispersive focal plane S4. The detector array comprised two multi-wire drift chambers (MWDCs) for trajectory reconstruction and four plastic scintillation detectors (SC and PL41-PL43) for triggering, time-of-flight, and energy-loss measurements.}
\label{fig:FRS}
\end{figure}

The experiment was performed at GSI Helmholtzzentrum für Schwerionenforschung in Darmstadt, Germany, in conjunction with the studies reported in Refs.~\cite{Tanaka2016,Tanaka2018}. Proton beams from the SIS-18 synchrotron at incident energies of 403, 604, 907 and 1209 MeV, with a typical intensity of about $10^{10}$ particles per spill and a spot size of approximately $2\times3$ mm$^2$, struck  a polyoxymethylene (POM; (CH$_2$O)$_n$) strip target with a horizontal width of 1 mm placed at the target station of the Fragment Separator (FRS)~\cite{Geissel1992} (see Fig.~\ref{fig:FRS}(a)). The POM target served as the oxygen target, while  additional carbon (C) and deuterated polyethylene (CD$_2$) targets were used for background subtraction and transmission-efficiency calibration, respectively. Hydrogen in POM does not contribute to deuteron production via the reaction of interest. 
The CD$_2$ target provided a continuous momentum distribution from $p$+$d$ elastic scattering for determining the FRS transmission efficiency. Target thicknesses and corresponding incident beam energies are listed in Table~\ref{tab:target}.

\begin{table}[t]
\centering
\caption{Targets used in the experiment. $E_{\textrm{inc}}$ denotes the incident-proton energy, and the last column lists the reactions analyzed for each target.}
\begin{tabular}{c c c c}
\hline \hline
Target & Thickness & $E_{\textrm{inc}}$ & Reactions \\
 & [mg/cm$^2$] & [MeV] & \\ 
\hline
POM     & 100$\pm$3     & 403, 604    & $^{16}$O, $^{12}$C$(p,d)$\\
C       & 107$\pm$4     & 403, 604    & $^{12}$C$(p,d)$\\
POM     & 308$\pm$6     & 907, 1209   & $^{16}$O, $^{12}$C$(p,d)$\\
C       & 327$\pm1$     & 907, 1209   & $^{12}$C$(p,d)$\\
CD$_2$  & 1027$\pm$2    & 403           & $^2$H$(p,d)$\\
\hline\hline 
\end{tabular}
\label{tab:target}
\end{table}

The FRS was operated as a zero-degree spectrometer in its high-resolution, four-stage dispersive configuration~\cite{Haettner2020}, providing a momentum dispersion of about 19 cm/\% at the final focal plane (S4), where deuterons were detected. The detector setup at S4, shown in Fig.~\ref{fig:FRS}(b), comprised two multi-wire drift chambers (MWDCs) for trajectory reconstruction and four plastic scintillation detectors: SC for generating the trigger singal and measuring the energy loss ($\Delta E$), and PL41-PL43 for time-of-flight (TOF) and $\Delta E$ measurements used in particle identification. 
Each MWDC featured eight wire planes covering an active area of $240\times140$ mm$^2$, with a position resolution of approximately 100 $\mu$m. 

Multiple magnetic-rigidity settings were used at 403 MeV to cover the states of interest, whereas a single setting sufficed for both carbon and POM targets at higher energies. Background contributions from carbon were subtracted using dedicated reference measurements, and the final excitation-energy spectra were corrected for FRS transmission efficiency. 
The deuteron momentum was reconstructed from the position at S4. The nonuniform momentum acceptance was measured with the 1027-mg/cm$^2$ CD$_2$ target at three central-momentum settings and used to correct the yields.
Further details on the experimental conditions and detector setup are provided in Refs.~\cite{Tanaka2016,Tanaka2018}.

\section{Data Analysis} 
\label{sec:data_analysis}

\subsection{Particle identification} 
\label{sub:particle_identification}

\begin{figure}
\includegraphics[width=\columnwidth]{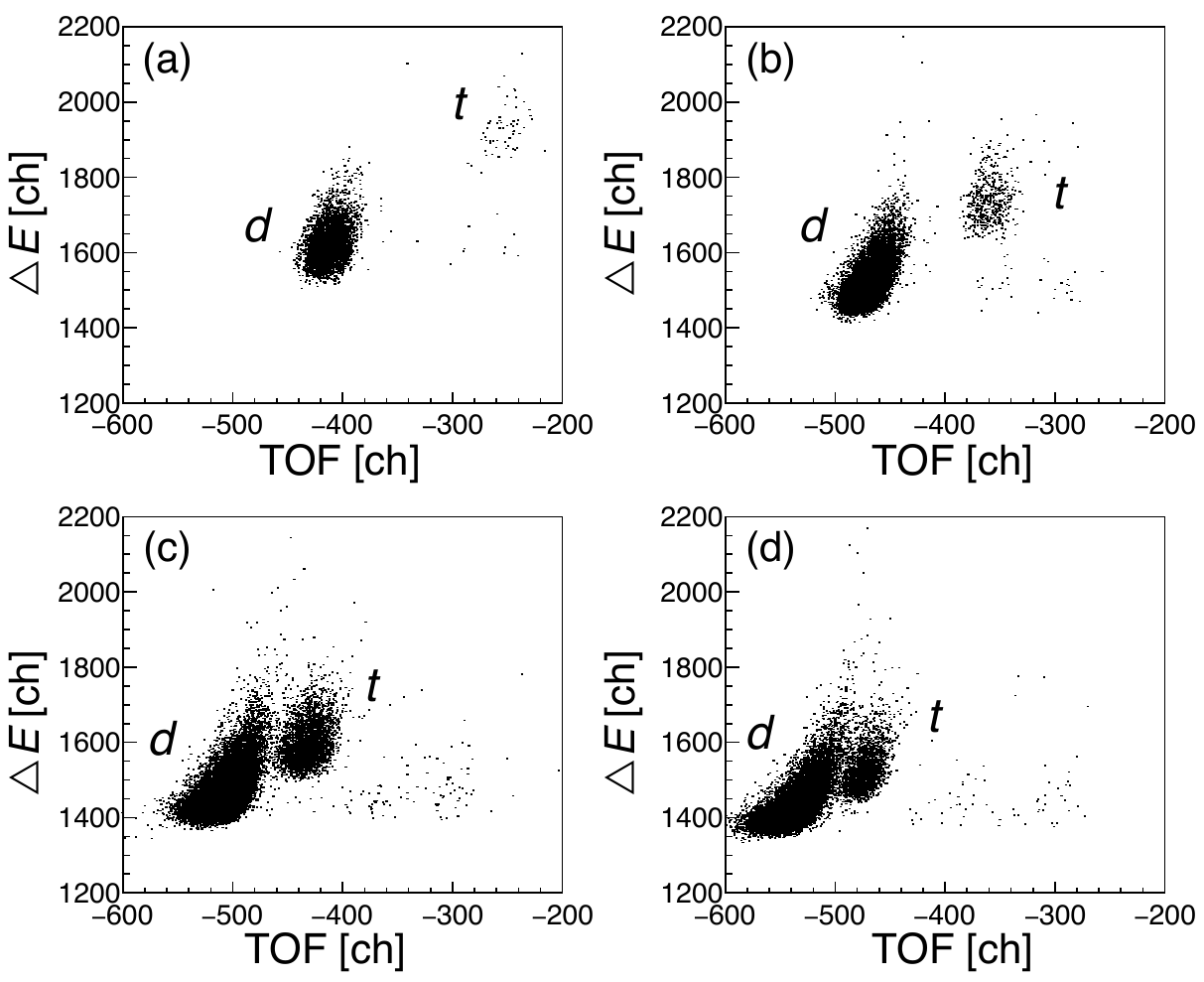}%
\caption{ Particle identification at S4 using the $\Delta E-$TOF method.  The horizontal axis represents the time difference between SC and PL42/PL43 detectors, and the vertical axis shows the energy deposited in SC.  Panels (a)--(d) correspond to incident proton energies of 403, 604, 907 and 1209 MeV, respectively.}
\label{fig:PID}
\end{figure}

Particle identification was achieved using the $\Delta E-$TOF measurement at S4, as shown in Fig. \ref{fig:PID}.  
The magnetic rigidity of deuterons emitted in the $(p,d)$ reactions exceeded that of the incident protons by more than 25\%, making the proton background negligible. 
The dominant background at the final focal plane consisted of tritons, which may originate from the target or elsewhere in the FRS. As clearly seen in Fig.~\ref{fig:PID}, tritons are well separated from deuterons at all beam energies. 
Furthermore, the triton momentum spectrum was confirmed to be broadly distributed and exhibiting no peak structures. Consequently, the momentum spectra obtained after particle identification are purely attributable to deuterons.

\subsection{Momentum distribution} 
\label{sub:momentum_distribution}

\begin{figure}
\includegraphics[width=\columnwidth]{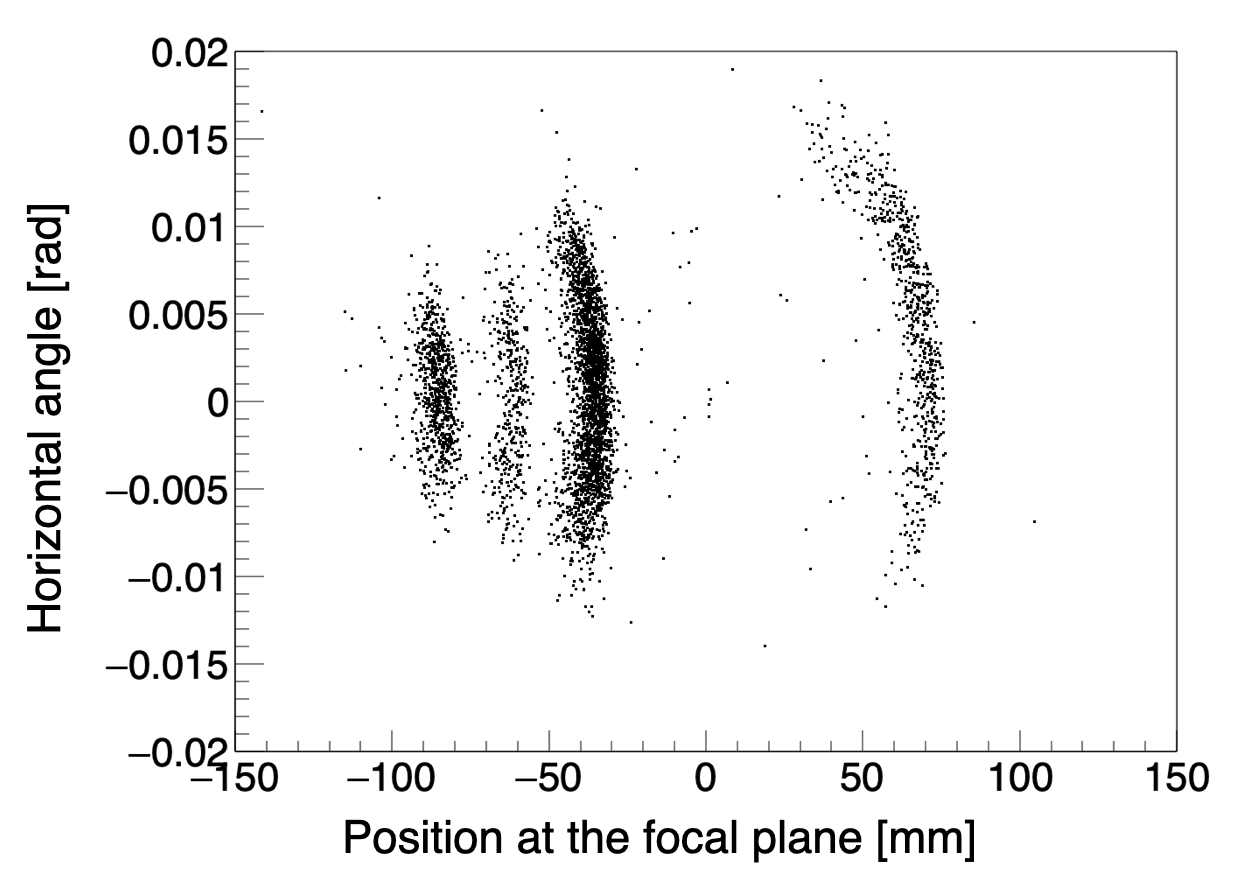}%
\caption{Horizontal position and angle distributions of deuterons at the S4 focal plane. The correlation between position and angle is used to correct for optical aberrations and improve the momentum resolution.}
\label{fig:AX}
\end{figure}

\begin{figure}
\includegraphics[width=\columnwidth]{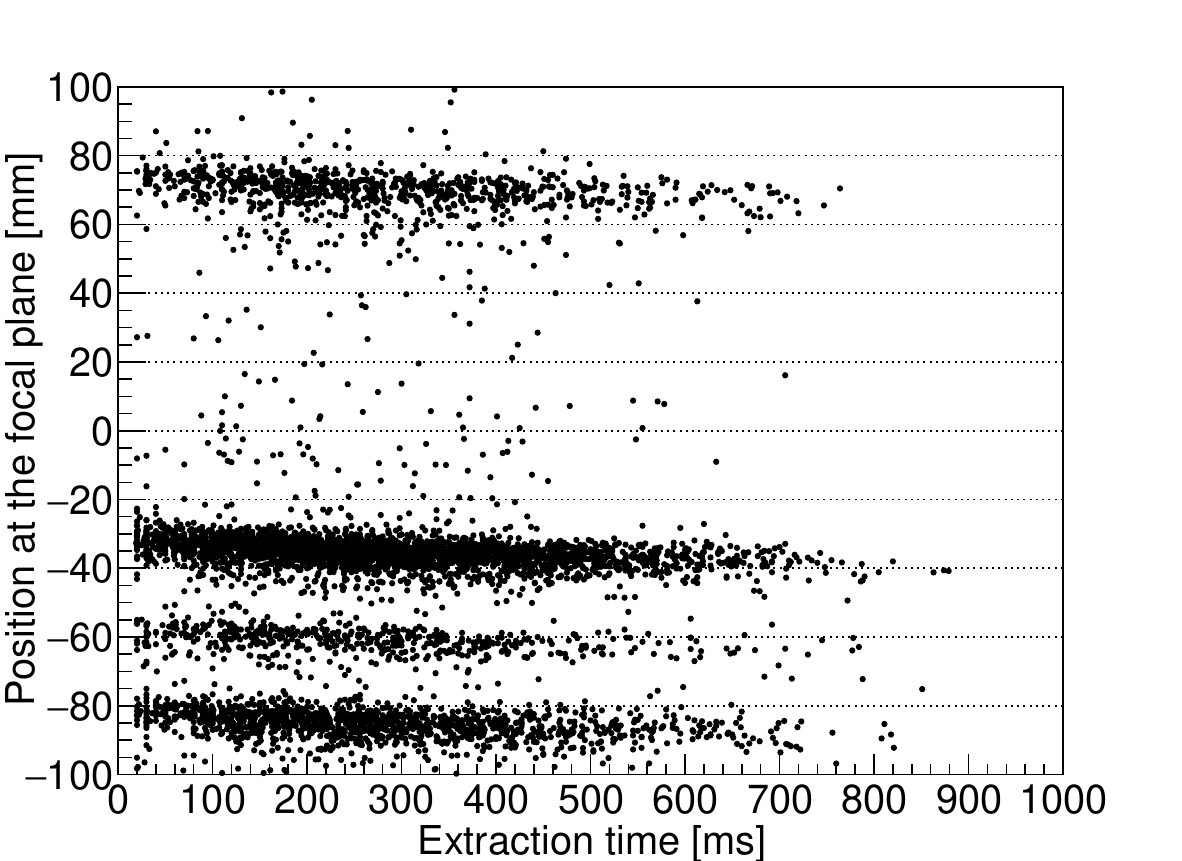}%
\caption{Particle position at the focal plane as a function of time since the start of beam extraction. The gradual shift reflects the energy drift of the incident proton beam.}
\label{fig:ext}
\end{figure}

\begin{figure}
\includegraphics[width=\columnwidth]{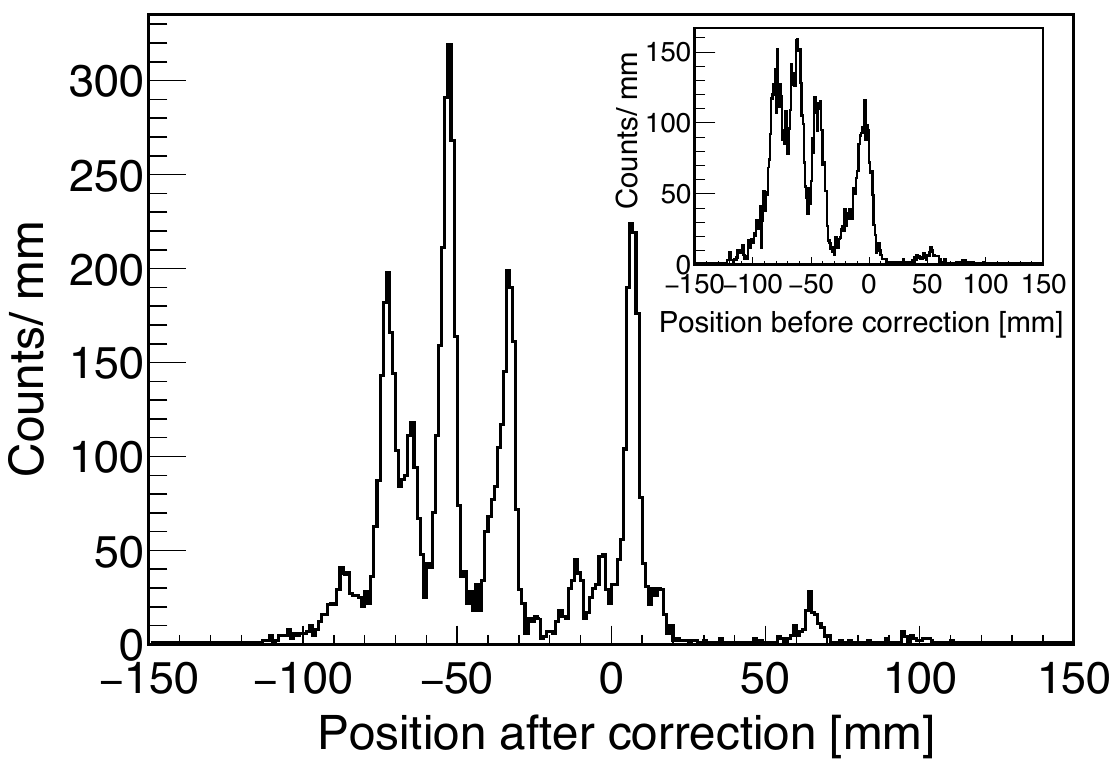}%
\caption{Position spectrum of deuterons emitted from the POM target at an incident proton energy of 1209 MeV, after applying the three corrections described in the text. The inset shows the spectrum before the corrections, illustrating the significant improvement in resolution achieved.}
\label{fig:X4}
\end{figure}

The momentum of each deuteron was determined from its reconstructed position at the focal plane, which is tilted by 9$^{\circ}$ relative to the beam line.  
Figure~\ref{fig:AX} shows an example of the particle position distribution at the focal plane for a POM target run, where different final states appear as distinct event clusters.

Due to optical aberrations, particles with the same momentum but different angles are not focused to the same position. To obtain high-resolution momentum spectra, the following corrections were applied, based on the assumption that each final state produces a narrow horizontal-width cluster and that deuterons from different states form lines at specific horizontal positions:
\begin{enumerate}
\item Higher-order optical aberrations were corrected to eliminate the angular dependence of the observed positions.
\item The incident beam energy was found to vary with extraction time. Figure~\ref{fig:ext} shows the detected particle position as a function of the time elapsed since the start of beam extraction. All position data shift consistently with time, indicating a drift in the incident beam energy. This drift was corrected by assuming a linear dependence of the beam energy on extraction time.
\item Small time-dependent shifts in particle positions were also observed and corrected by monitoring the position spectrum as a function of detection time.
\end{enumerate}
Figure~\ref{fig:X4} presents an example of the position spectrum after applying these three corrections, with the uncorrected spectrum shown in the inset for comparison. The improvement in resolution is clearly visible.

The deuteron momentum $k_d$ was determined from the position $x$ using the linear relation
\beq
k_d = (\frac{x}{D}+1)k_0
\eeq
where $k_0$ is the central momentum setting of the FRS, and $D$ is the momentum dispersion at the focal plane. The consistency of the observed deuteron momenta was confirmed using the known masses of all involved particles and the excitation energies of the $^{15}$O states.

\subsection{Transmission efficiency} 
\label{sub:transmission_efficiency}

\begin{figure}
\includegraphics[width=\columnwidth]{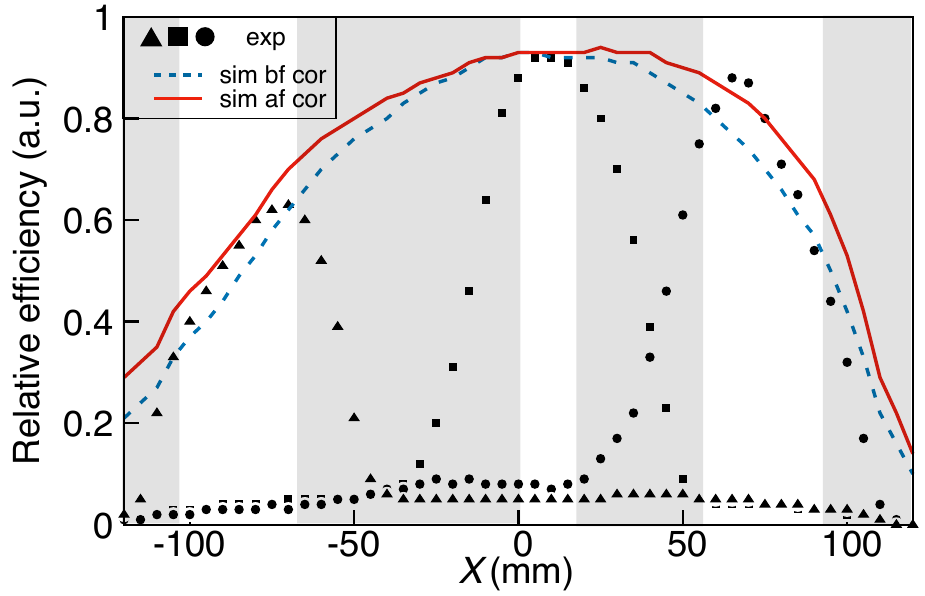}%
\caption{Transmission efficiency of the FRS in the high-resolution mode at the final focal plane (S4). Black symbols represent the experimentally determined transmission efficiency obtained from $p$+$d$ elastic scattering measurements at three different momentum settings. The blue-dashed curve is the Monte Carlo simulation result, while the red-solid curve is the result after the phenomenological correction described in the text. The grey-shaded regions indicate areas excluded from the comparison due to nonuniform deuteron distributions.}
\label{fig:tran}
\end{figure}

The transmission efficiency from the target to the S4 focal plane was simulated using Monte Carlo code MOCADI~\cite{Iwasa1997}, with the FRS transfer matrices provided by the GICOSY code~\cite{Makino2006}. The simulation was validated using the deuteron momentum distribution from $p$+$d$ elastic scattering at 403 MeV. A thick CD$_2$ target was employed to produce a nearly uniform deuteron spectrum over a limited momentum range; however, a single magnetic-rigidity setting was insufficient to cover the full momentum acceptance. Therefore, measurements were performed at three different central-momentum settings to map the entire acceptance region.

The measured relative transmission efficiency is shown by the black symbols in Fig.~\ref{fig:tran}, and the simulated efficiency is shown by the blue dashed curve. The simulation deviates from the data outside the grey-shaded regions.  To better reproduce the measured efficiency, a phenomenological quadratic correction was applied to the simulated efficiency, yielding the red solid curve in the figure. The difference between the corrected and original simulated efficiencies was taken as the systematic uncertainty. Within the flat region of the momentum spectra, the agreement between the simulation and the measurement is satisfactory. The same simulation and correction procedure was applied consistently at all other incident energies.

\subsection{ Reconstruction of excitation-energy spectra and determination of cross-section ratios} 
\label{sub:determination_of_cross_section_ratios}

The final states in oxygen and carbon were assigned based on the calculated deuteron momenta and the known excitation energies. At 403 MeV, the carbon and POM data were taken with different magnetic-rigidity settings, and overlapping contributions were treated via peak-area subtraction. For incident energies of 604, 907 and 1209 MeV, the carbon spectra measured with the corresponding rigidity settings were incorporated into a global fit of the POM spectra, following the procedures described in Ref.~\cite{Wang2026}. In this approach, the relative amplitudes of the $^{11}$C peaks were fixed from the carbon-target fit, while the overall carbon normalization and the oxygen-state amplitudes were determined from the POM data.

The carbon-target position distributions were used to determine the carbon contribution to the POM spectra. Their excitation energies and peak assignments are consistent with the known low-lying states of $^{11}$C. The peak shapes and relative amplitudes were obtained from a fitting function comprising four peaks:
\beq
&&y=y_0+\sum^{4}_{i=1}A_i f_i(x) ~,
\eeq
where $y_0$ is a constant background, and $A_i$ is the amplitude of the $i$-th peak. Because the spectral peaks are asymmetric, each was described by a double-Gaussian function:
\begin{equation}
\begin{split}
f_i(x)={}&\frac{1-a}{\sqrt{2\pi}\sigma_1}
 e^{-\frac{1}{2}\left(\frac{x-x_i}{\sigma_1}\right)^2}\\
 &+\frac{a}{\sqrt{2\pi}\sigma_2}
 e^{-\frac{1}{2}\left(\frac{x-x_i-d}{\sigma_2}\right)^2},
\end{split}
\label{eq:doub_gaus}
\end{equation}
where $a,~\sigma_1,~ \sigma_2$ and $d$ are free shape parameters common to all states.
Only the ground state (3/2$^-$) and three low-lying excited states of $^{11}$C -- the 1/2$^-$ state at 2.0 MeV, 5/2$^-$ state at 4.3 MeV, and 3/2$^-$ state at 4.8 MeV -- were included in the fit. No other peaks were observed in the carbon-target data within the energy region of interest. Including additional higher-lying states was found to have negligible effect on the amplitudes of the four low-lying peaks used for the subtraction.

\label{ssub:O_spectra}

\begin{figure*}
\includegraphics[width=1.0\textwidth]{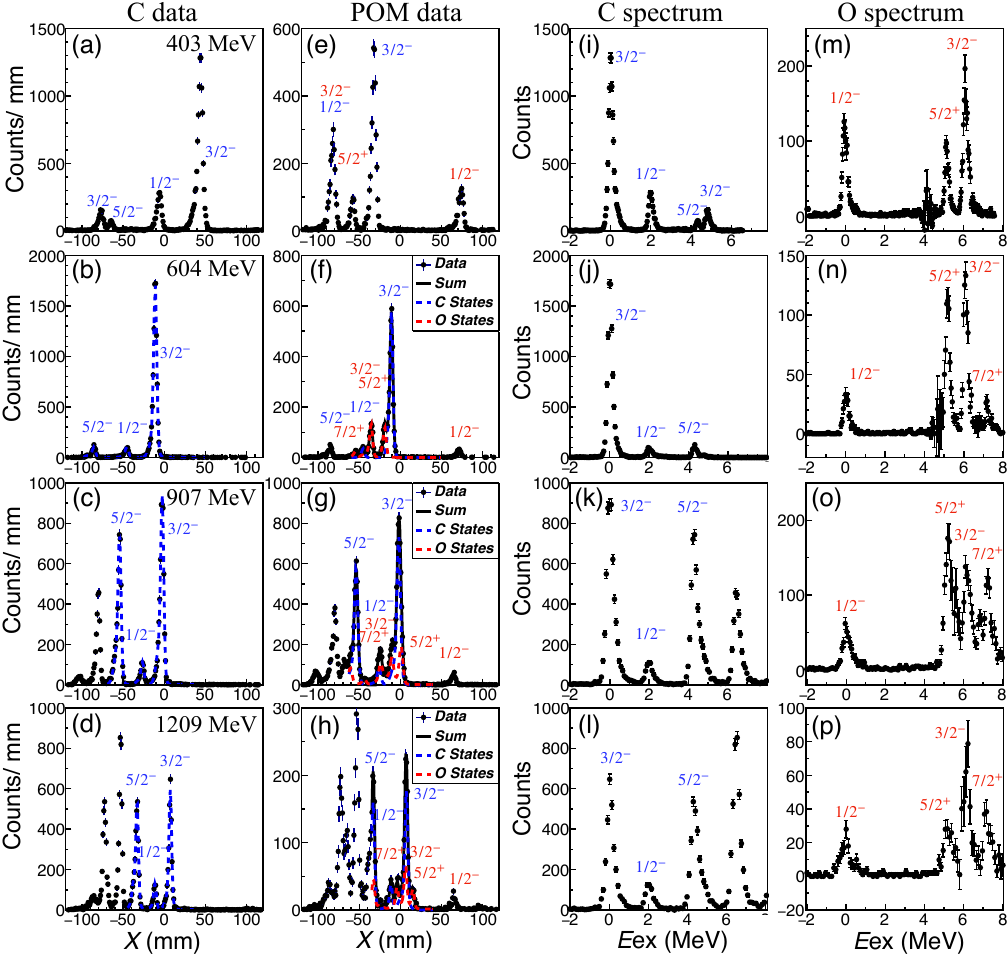}%
\caption{Position and excitation-energy spectra used in the carbon-subtraction procedure. The columns from left to right show: C-target position distributions (a)-(d), POM-target position distributions (e)-(h), $^{11}$C excitation-energy spectra (i)-(l), and $^{15}$O excitation-energy spectra obtained after carbon subtraction (m)-(p). The rows correspond to incident-proton energies of 403 MeV: (a), (e), (i) and (m); 604 MeV: (b), (f), (j) and (n); 907 MeV: (c), (g), (k) and (o); and 1209 MeV: (d), (h), (l) and (p). In the POM position spectra, the black-solid curves represent the total fits, while the blue- and red-dashed curves indicate the C and O contributions, respectively. Spin-parities for the selected key states in $^{11}$C and $^{15}$O are labeled in blue and red, respectively.}
\label{fig:specPOM}
\end{figure*}

The position distributions and corresponding excitation-energy spectra for the C and POM targets are presented in Fig.~\ref{fig:specPOM}. The carbon contribution is constrained by the carbon-target fit after correcting for the difference in target thickness, while the oxygen amplitudes are left free in the global fit. The same double-Gaussian peak shape is used for both carbon and oxygen components, following Ref.~\cite{Wang2026}. 
The fit includes the known $^{15}$O states at 5.24 (5/2$^+$), 6.18 (3/2$^-$), 6.79 (3/2$^+$), 6.86 (5/2$^+$), 7.28 (7/2$^+$), 8.92 (5/2$^+$), 9.66 (3/2$^-$), and 10.30 MeV (5/2$^+$).
The 5.2-MeV peak was fitted at the energy of the 5.24-MeV 5/2$^+$ state; the contribution of the 5.18-MeV 1/2$^+$ state is expected to be negligible at the high incident energies as mentioned in Ref.~\cite{Wang2026}. Consequently, $R_{5.2}$ is interpreted predominantly as the ratio for the 5/2$^+$ state.

The fitted oxygen spectra are shown by red dashed curves, and the final total fits to the full spectra are shown by black solid curves. Subtracting the fitted carbon spectra from the POM spectra yields the  oxygen spectra shown in the rightmost column [panels (m)--(p)]. As seen in all spectra, the background level is much smaller than the true counts ($<5\%$).  
The counts for each excited state of $^{15}$O were integrated and corrected for FRS transmission efficiency to obtain the cross-section ratios relative to the ground-state transition.

At 403 MeV, the carbon and POM measurements were performed with different magnetic-rigidity settings. In this case, the 6.18-MeV ($3/2^-$) state in $^{15}$O and 2.00-MeV ($1/2^-$) state in $^{11}$C overlap, whereas the other relevant low-lying peaks are well separated. Subtraction was therefore performed using integrated peak areas rather than the global spectral fit used at the higher energies. After correcting for the carbon-target normalization and FRS transmission efficiencies, the $^{11}$C 2.00-MeV area was subtracted from the overlapping POM peak. 

The transmission-efficiency-corrected count ratios for transitions to the various excited states of $^{15}$O, normalized to the ground-state transition, are tabulated in Table~\ref{tab:ratio}. The ratios are defined as follows:
\beq
&&R_{5.2} = \frac{\sigma(^{15}\textrm{O}_{5.2})}{\sigma(^{15}\textrm{O}_{\textrm{g.s.}})}, \\
&&R_{6.2} = \frac{\sigma(^{15}\textrm{O}_{6.2})}{\sigma(^{15}\textrm{O}_{\textrm{g.s.}})}, \\
&&R_{7.3} = \frac{\sigma(^{15}\textrm{O}_{7.3})}{\sigma(^{15}\textrm{O}_{\textrm{g.s.}})}, 
\eeq
where $\sigma (^{15}\text{O}_x)$ denotes the cross section for the transition to the final state $x$.
The ratio $R_{7.3}$ could not be extracted at 403 MeV because the 7.28-MeV state fell outside the limited momentum acceptance.

For each incident energy, the momentum transfer $q$ was calculated separately for the two states entering a given ratio; the average of these two values was adopted as the momentum transfer for that ratio, with half their difference taken as the associated uncertainty.
The quoted uncertainties on the ratios include both statistical and systematic contributions; the latter arise from the estimation of the transmission efficiency, the choice of fitting functions, and the analysis procedures, with typical values given in Ref.~\cite{Wang2026}.

\begin{table*} 
\centering
\caption{Cross-section ratios $R_{5.2}$, $R_{6.2}$, and $R_{7.3}$ for various average momentum transfers $q$ corresponding to transitions to the excited and ground states. The value of $q$ for each ratio is the mean of the momentum transfers for the two involved states, with uncertainty taken as half their difference. Ratio uncertainties include statistical (sta), systematic from transmission efficiency (sys1), and systematic from analysis procedures (sys2). The total uncertainty (tot) is the quadratic sum of the statistical uncertainty and the larger of the two systematic uncertainties. Values for $R_{5.2}$ and $R_{6.2}$ are consistent with those reported in Ref.~\cite{Wang2026}; $R_{7.3}$ is reported here for the first time.}
\small
\setlength{\tabcolsep}{2.75pt}
\begin{tabular}{c || c | c c c c c || c | c c c c c || c | c c c c c}
\hline\hline
Energy & $q$ & \multirow{2}{*}{$R_{5.2}$} & \multicolumn{4}{c||}{Error} & $q$ & \multirow{2}{*}{$R_{6.2}$} & \multicolumn{4}{c||}{Error} & $q$ & \multirow{2}{*}{$R_{7.3}$} & \multicolumn{4}{c}{Error} \\
(MeV) & [fm$^{-1}$] & & sta & sys1 & sys2 & tot & [fm$^{-1}$] &  & sta & sys1 & sys2 & tot & [fm$^{-1}$] &  & sta & sys1 & sys2 & tot \\
\hline 
403  & 1.53(2) & 0.60 & 0.02 & 0.06  & -   & 0.07 & 1.52(3) & 1.87 & 0.05 & 0.19  & -  & 0.19 & - & - & - & - & - & - \\
604  & 1.85(2) & 3.2 & 0.3 & 0.3 & 0.7 & 0.7 & 1.85(2) & 3.63 & 0.30 & 0.30 & 0.08 & 0.42 & 1.85(3) & 0.66 & 0.08 & 0.03 & 0.02 & 0.09 \\
907  & 2.21(2) & 3.1 & 0.3 & 0.3 & 0.2  & 0.4 & 2.20(2) & 1.99 & 0.30 & 0.19 & 0.16 & 0.36 & 2.20(2) & 1.60 & 0.14 & 0.13 & 0.02 & 0.19 \\
1209 & 2.47(2) & 1.22 & 0.17 & 0.12 & 0.09 & 0.20 & 2.46(2) & 2.7 & 0.6 & 0.3 & 0.5 & 0.8 & 2.46(2) & 0.98 & 0.17 & 0.08 & 0.32 & 0.37 \\
\hline\hline 
\end{tabular}
\label{tab:ratio}
\end{table*}

\section{Results and Discussion} 
\label{sec:results_and_discussion}

\begin{figure}
\includegraphics[width=\columnwidth]{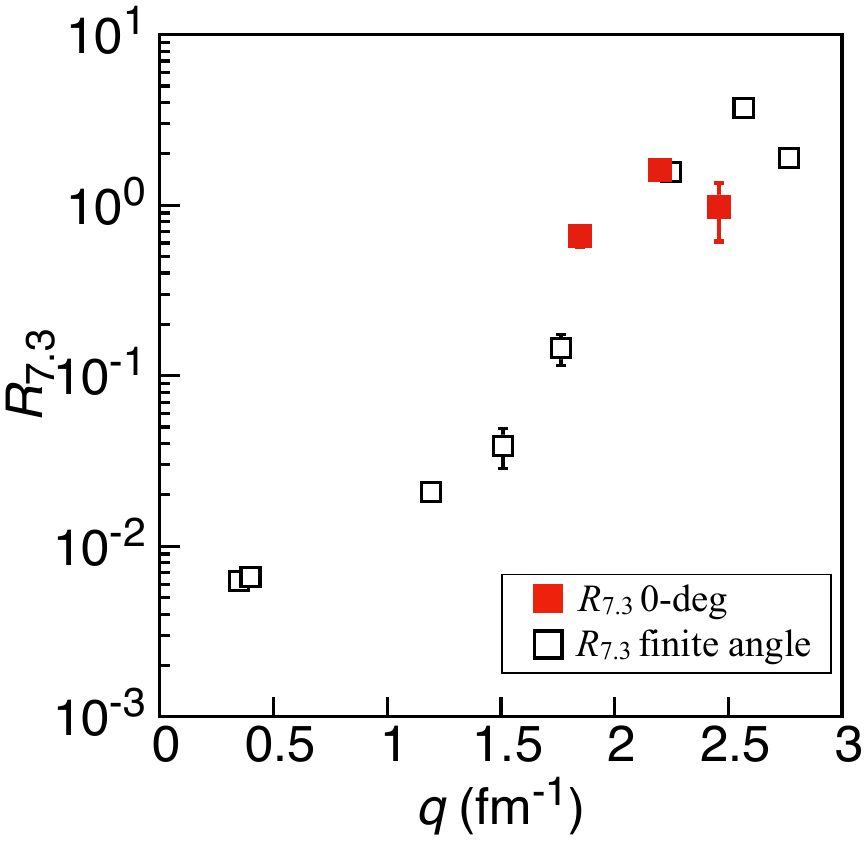}%
\caption{Cross-section ratio $R_{7.3}$ as a function of momentum transfer $q$. Red-filled squares represent the present zero-degree data, with error bars indicating combined statistical and systematic uncertainties. Open symbols show previous finite-angle measurements at $\theta_{d}<15^\circ$~\cite{Snelgrove1969,Ong2013,Smith1984}.}
\label{fig:Rplot}
\end{figure}

\begin{figure}
\includegraphics[width=\columnwidth]{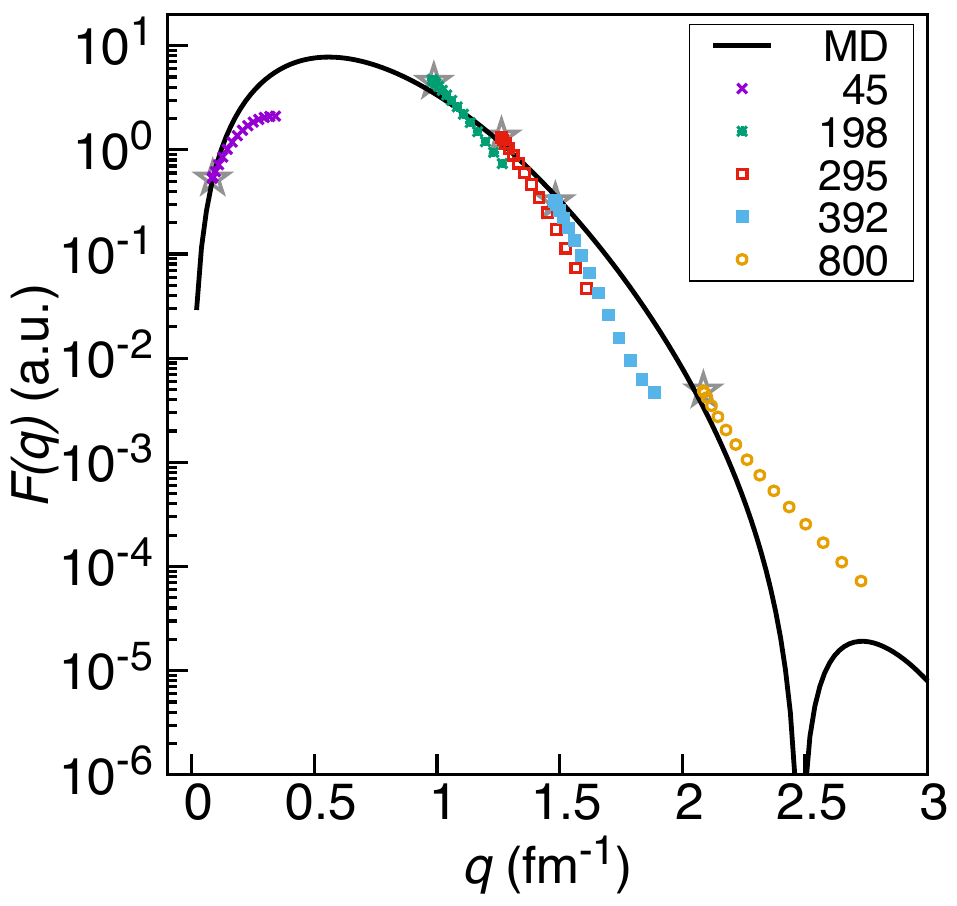}%
\caption{Calculated cross sections for the negative-parity 6.18-MeV 3/2$^-$ transition as a function of momentum transfer. Symbols represent ADWA results at the indicated incident energies, with zero-degree points marked by stars.  The solid curve shows the PWBA/momentum-distribution (MD) dependence, scaled to match the 45-MeV zero-degree result.}
\label{fig:DWBA}
\end{figure}

The momentum dependence of $R_{6.2}$ and $R_{5.2}$ was previously reported in Ref.~\cite{Wang2026}. In the present work, we extend this analysis by presenting the momentum dependence of $R_{7.3}$, which corresponds to the positive-parity 7.28-MeV 7/2$^+$ state. The complete set of zero-degree ratios is compiled in Table~\ref{tab:ratio}.
Figure~\ref{fig:Rplot} presents the newly determined momentum dependence of $R_{7.3}$, together with earlier measurements taken at finite angles (center-of-mass deuteron angles below $15^\circ$) and incident-proton energies of 45~MeV~\cite{Snelgrove1969}, 198--392~MeV~\cite{Ong2013}, and 800~MeV~\cite{Smith1984}. For the 800-MeV data, values read directly from the published figure are used, as they differ from those tabulated in EXFOR~\cite{EXFOR2018}.

The ratio $R_{7.3}$ exhibits a steep rise of more than two orders of magnitude from the lowest measured momentum transfer to $q=2$~fm$^{-1}$, resembling the behavior of $R_{5.2}$ and suggesting enhanced high-momentum strength in transition to these positive-parity states. In contrast, $R_{6.2}$ shows only a modest increase, remaining within one order of magnitude. Together with the results reported in Ref.~\cite{Wang2026}, the present measurements extend these systematics to cover a more complete momentum-transfer region and establish the first zero-degree determination of $R_{7.3}$.

To interpret these trends in terms of nuclear structure, it is essential to establish that the measured cross-section ratios reliably reflect the underlying momentum distributions~\cite{Tropiano2021}. While the quantitative interpretation of absolute high-energy ($p,d$) cross sections is subject to established theoretical uncertainties -- such as relativistic corrections~\cite{Kamada2026,Hai_submit}, optical-potential ambiguities~\cite{Ramon2026}, and reaction-model dependence~\cite{Shim2026} -- the kinematic conditions at zero degrees offer a viable pathway to constrain nuclear structure. Specifically, as shown below, under zero-degree scattering conditions, adiabatic distorted wave approximation (ADWA) calculations converge with the plane-wave Born approximation (PWBA), where the differential cross section as a function of momentum transfer becomes directly proportional to the nucleon momentum distribution within the nucleus.

This kinematic equivalence at zero degrees implies that systematic distortions affecting the absolute magnitude largely cancel in relative observables. Consequently, it is reasonable to directly compare experimental cross-section ratios measured at zero degrees with momentum-distribution ratios derived from {\it ab initio} nuclear structure theory, thereby isolating the underlying single-particle dynamics from reaction-mechanism artifacts. 
The influence of distortion was examined with the ADWA accounting for relativistic kinematic corrections~\cite{Hai_submit} for the negative-parity 6.18-MeV $3/2^-$ transition, as shown in Fig.~\ref{fig:DWBA}. The calculated distributions deviate at finite angles, whereas the zero-degree points remain close to the PWBA/momentum-distribution (MD) dependence. Furthermore, the ground-state and 6.18-MeV transitions also have nearly identical high-energy kinematics, so much of the remaining distortion is expected to cancel in their cross-section ratio. The resulting agreement supports the use of the zero-degree ratio to investigate relative momentum trends.

It should be noted, however, that the same calculation cannot be used to interpret the positive-parity angular distributions quantitatively. The adopted positive-parity bound-state wave functions lack the high-momentum components central to the transitions under study, and previous calculations using coupled-channels Born approximation (CCBA)~\cite{Smith1984} and continuum-discretized coupled-channel method with Born approximation (CDCC-BA)~\cite{Ong2013} accordingly fail to reproduce their absolute differential cross sections at high energy. This limitation does not invalidate the measured ratios: $R_{5.2}$ and $R_{7.3}$ remain valid for comparing relative evolution with momentum transfer, even if an absolute theoretical description of the positive-parity strength is not claimed.

Assuming the zero-degree ratio provides a reliable measure of the relative momentum dependence, we consider the origin of the observed positive-parity strength. Tensor-induced high-momentum components are expected to arise from the mixing of 2p-2h or higher excitations with $\Delta L=2,~\Delta S=2$. Such excitations cannot occur entirely within the $p$-shell, which constitutes the dominant configuration of the $^{16}$O ground state. Instead, the relevant orbital mixing involves promotion of nucleons to higher major shells, such as the $sd$-shell. Single-particle levels in the $sd$-shell carry positive parity, in contrast to the negative-parity $p$-shell orbitals. This parity difference enables the identification of whether the picked-up neutron originates from the $p$-shell or the $sd$-shell based on the parity of the final state in $^{15}$O.

The $^{15}$O ground state ($1/2^-$) and the 6.18-MeV excited state ($3/2^-$) have negative parity and are understood to arise predominantly from neutron pickup from the $p_{1/2}$ and $p_{3/2}$ orbitals, respectively. These transitions are not expected to be significantly influenced by tensor correlations. As mentioned earlier (see Sec.~\ref{ssub:O_spectra}), the 5.2-MeV peak is dominated by the 5.24-MeV ($5/2^+$) state, which carries positive parity and can arise from pickup from the $d_{5/2}$ orbital, making it sensitive to tensor-induced effects. The 7.3-MeV ($7/2^+$) state also carries positive parity. Although a simple 2p-2h configuration would suggest excitation to the $g_{7/2}$-shell (two major shells above the $sd$-shell), it may instead arise from more complicated mechanisms such as multi-particle-multi-hole configurations.

\begin{figure}
\includegraphics[width=\columnwidth]{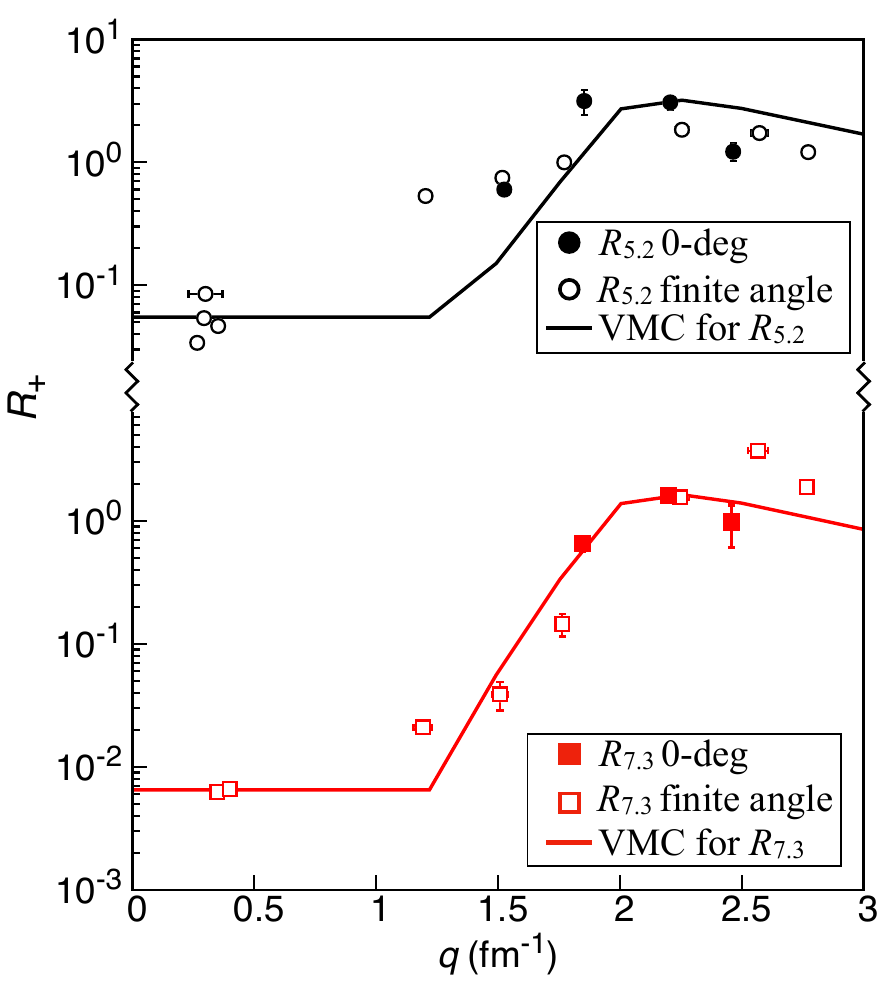}%
\caption{Positive-parity ratios $R_{5.2}$ and $R_{7.3}$ compared with representative VMC trends. Symbols denote the experimental data, while curves show the relative momentum dependence calculated using the prescription of Ref.~\cite{Wang2026}.}
\label{fig:discu}
\end{figure}

In Ref.~\cite{Wang2026}, the measured cross-section ratios were compared with several theoretical calculations~\cite{Pieper1992,Neff2003,Alvioli2008,Fabrocini2001} via $R_{+}(q) = u+v f(q)$, where $u$ and $v$ are free parameters that depend on the states involved, and $f(q)$ is the momentum-distribution ratio obtained from the calculations. All models give trends qualitatively consistent with the measurement, so for clarity we show only Variational Monte Carlo (VMC) calculations~\cite{Pieper1992} as a representative example in Fig.~\ref{fig:discu}. 
The black curve employs mixing parameters $u=0.055$ and $v=0.29$ for $R_{5.2}$, and the red curve employs $u=0.0065$ and $v=0.15$ for $R_{7.3}$. It is important to note that these curves do not represent orbital-by-orbital calculations of the specific positive-parity states; therefore, the comparison is restricted to the relative momentum dependence of the measured ratios.
Figure~\ref{fig:discu} shows that both $R_{5.2}$ and $R_{7.3}$ rise strongly toward intermediate momentum transfer and follow the broad VMC trend. 
A peak structure is observed around 2 fm$^{-1}$ for $R_{7.3}$, similar to that for $R_{5.2}$. Differences between the two measured ratios and the VMC curves therefore motivate structure calculations that provide realistic high-momentum components separately for the $5/2^+$ and $7/2^+$ states. Additional zero-degree measurements, especially at lower momentum transfer, would further reduce the ambiguity associated with the finite-angle data.

\section{Summary} 
\label{sec:summary}

We have presented a comprehensive experimental account of the zero-degree $^{16}$O$(p,d)^{15}$O measurements at incident energies of 403, 604, 907, and 1209 MeV. The present work extends the study reported in Ref.~\cite{Wang2026} by providing additional experimental and data-analysis details, and by reporting for the first time the ratio $R_{7.3}$, defined as the cross section populating the 7.28-MeV ($7/2^+$) state relative to that of the ground state. Detailed descriptions of the detector setup, momentum reconstruction, FRS transmission-efficiency corrections, and carbon-background subtraction complement the principal physics results of the previous work. 

The measured cross-section ratios include the negative-parity reference $R_{6.2}$ and the positive-parity ratios $R_{5.2}$ and $R_{7.3}$. ADWA calculations demonstrate that the zero-degree negative-parity results are least affected by distortion effects and closely follow the momentum-distribution dependence, thereby supporting the ratio comparison. For the positive-parity states, standard wave functions lacking realistic high-momentum components fail to reproduce the absolute angular distributions; consequently, the data are interpreted through their relative momentum dependence. Both $R_{5.2}$ and $R_{7.3}$ exhibit a characteristic enhancement toward intermediate momentum transfer, with a pronounced peak structure around 2 fm$^{-1}$, consistent with  nuclear models that include the effects of tensor interactions.

\section*{Statements and Declarations}
\paragraph{Funding} This work was supported by the Japan Society for the Promotion of Science Grant-in-Aid for Scientific Research under the contract number 23224008, the Hessian Ministry for Science and Art (HMWK) through the LOEWE Center HICforFAIR, Justus-Liebig-Universität Gießen and GSI under the JLU-GSI strategic Helmholtz partnership agreement. It was also supported by National Natural Science Foundation of China (Nos. 12175009, 12175280), the International Partnership Program of the Chinese Academy of Sciences (016GJHZ2023063GC), and Major Science and Technology Projects in Gansu Province (24ZD13GA005).

\paragraph{Competing interests} The authors declare no competing interests.

\paragraph{Data availability} The cross-section ratios and their uncertainties are reported in Table~\ref{tab:ratio}.

\bibliographystyle{unsrt}
\bibliography{s436paper}
\end{document}